\documentclass[onecolumn,noshowpacs,nofootinbib,colorlinks,hyperindex,unicode,11pt]{revtex4}
\usepackage{graphicx}
\usepackage{tikz}

\usepackage[compat=1.1.0]{tikz-feynman}

\usepackage{amsfonts,amsmath,amssymb,tikz,accents}
\usepackage[eulergreek]{sansmath}
\usepackage{indentfirst,bigints}
\usepackage{hyperref,upgreek}
\usepackage{bm,graphics,color}
\usepackage{feynmp-auto}
\usepackage{slashed}
\counterwithout{equation}{section}
\usepackage{hyperref}
\usepackage{pgfplots}
\pgfplotsset{compat=1.16}

\hypersetup{hidelinks,backref=true,pagebackref=true,hyperindex=true,colorlinks=true,breaklinks=true,urlcolor= blue}
\hypersetup{%
  colorlinks = true,
  linkcolor  = blue,
  citecolor = cyan,
}
\usepackage{textgreek}
\usepackage{color,xcolor}

\newcommand{\blt}{\textcolor{black}}

\def\0{\mbox{\boldmath$\displaystyle\mathbb{O}$}}

\def\p{\partial}

\newcommand\orcidroldao{{\href{https://orcid.org/0000-0003-3978-532X}{\orcidicon}}}
\newcommand{\orcidicon}{%
	\begin{tikzpicture}
	\draw[lime, fill=lime] (0,0)
		circle [radius=0.16]
		node[white] {{\fontfamily{qag}\selectfont \tiny ID}};
	\draw[white, fill=white] (-0.0625,0.095)
		circle [radius=0.007];
	\end{tikzpicture}	\hspace{-2mm}
}
\newcommand\orcidg{{\href{https://orcid.org/0000-0002-7942-7941}{\orcidicon}}}
\newcommand\orcidNog{{\href{https://orcid.org/0000-0002-2376-8253}{\orcidicon}}}

\usepackage{float,xcolor,upgreek}
\usepackage{color} 
\usepackage{tikz-feynman}
\tikzfeynmanset{compat=1.0.0}
\newcommand{\beq}{\begin{eqnarray}}
\newcommand{\eeq}{\end{eqnarray}}
\newcommand{\bea}{\begin{eqnarray}}
\newcommand{\eea}{\end{eqnarray}}
\DeclareMathOperator{\sign}{sign}

\begin{document}

\title{On  valley asymmetry in a topological interaction for quasi-particles}

\author{G. B. de Gracia\orcidg{}}
\affiliation{Federal University of ABC, Center of Mathematics,  Santo Andr\'e, 09210-580, Brazil.}
\email{g.gracia@ufabc.edu.br}
\author{B. M. Pimentel\orcidNog{}}
\affiliation{Institute of Theoretical Physics, Sao Paulo State University, 01156-970, S\~ao Paulo,  Brazil }
\email{bruto.max@unesp.br}
\author{R. da Rocha\orcidroldao\!\!}
\affiliation{Federal University of ABC, Center of Mathematics,  Santo Andr\'e 09210-580, Brazil.}
\email{roldao.rocha@ufabc.edu.br}


\begin{abstract}
\indent This paper is focused on investigating the effects of a statistical interaction for graphene-like systems, providing Haldane-like properties for topologically trivial lattices. The associated self-energy correction yields an effective next-nearest hopping, inducing the topological phase, whose specific solutions are scrutinized. In the case of an external magnetic field, it leads to a renormalized quasi-particle structure with generalized Landau levels and explicit valley asymmetry. 
 A suitable tool for implementing such achievements is a judicious indefinite metric quantization, leading to advances in field theory foundations. Since the topological behavior is encoded in the radiative corrections, an unequivocal treatment using an integral representation is carefully developed.
\end{abstract}

\maketitle


\section{Introduction}
\label{intro11}
 \indent There are several condensed matter systems whose low energy excitations can be described by a Dirac-like Lagrangian with Lorentz symmetry breaking term, associated with their specific drift velocities \cite{marl,shen}. The inclusion of non-linear interactions for this model can be conveniently investigated utilizing field theoretical tools. More specifically, some materials like graphene, states in HgTe/CdTe quantum wells, silicene, germanene, and transition metal dichalcogenides (TMD) have independent low-energy degrees of freedom called valleys. 
 Their manipulation in technological devices is associated with the field of valleytronics \cite{Schaibley}. 
 A valleytronic material system presents a band structure consisting of at least two degenerate valley states corresponding to local energy minima, which can be engendered to store and process information. 
 Regarding topological insulators, the spin plays a major role, being related to the field of spintronics \cite{Rajput}. \blt{Among other relevant models in statistical physics and condensed matter, the Haldane model plays a prominent role. It comprises a planar tight-binding model on a honeycomb lattice, wherein long-range interactions are described by a second nearest neighbor hopping term \cite{haldane}. The Haldane model has an intrinsic inter-site magnetic flux. In the low-energy regime, it presents quasi-particle excitations.}

 \blt{The second-order expansion around
Dirac points of its reciprocal lattice yields a momentum-dependent mass term \cite{tese, ger} that typically appears in a class of condensed matter models being associated with the possibility of a non-trivial topological phase} \cite{shen}. This is related to the fact that this mass term changes sign if one replaces the modulus of the spatial momentum from zero to a given critical limiting value. Regarding experimental achievements, the advances in material engineering enabled a realization of the Haldane model in Fe-based materials \cite{fe}. Recent investigations associated with valley filters \cite{f1}, using line defects for silicene \cite{sil} and one specific for graphene \cite{f2}, represent relevant tools to manipulate Dirac materials. We can also mention the theoretical formulation for a semimetal,  whose low-energy physics reproduces the Rarita--Schwinger model \cite{f3}. Ref. \cite{f4} provided a careful analysis of the phases and the topological structure of its associated quasi-particle regime.  \blt{Interestingly, these spin-3/2  fermionic excitations were already experimentally observed in  CoSi and RhSi} \cite{f5}. Several developments have been accomplished in the AdS/CMT setup, including holographic descriptions of topological semimetals \cite{Landsteiner:2015pdh}
 and topological insulators \cite{Murugan:2016zal}.\\
\indent   It is also relevant to mention the field of straintronics, comprising a complementary method for modifying properties in Dirac material \cite{strain}. Recent studies indicate that epitaxial strain can turn a standard metal into a superconductor \cite{strain2}. It means that its characteristic non-linear fermionic interaction can be induced by this method. Considering these advances, our main objective is to investigate a scenario compatible with discrete symmetry breaking, in which a statistical interaction emerges independently of its generating mechanism. It is possible to show that, with such an interaction, trivial Dirac materials can enter the same topological class as a given phase of the Haldane model. Self-energy corrections generate a momentum-dependent mass term with topological implications. It resembles the structure that naturally arises in \blt{the second-order expansion around
Dirac points of the reciprocal lattice in the Haldane model}.  In compliance with our investigations, it is worth mentioning that a hypothetical ordinary lattice insulator can become a topological one by a suitable inclusion of fermionic self-interactions \cite{bftop}.\\
\indent Another inspirational work is Ref. \cite{coulomb}, in which the effect of the Coulomb interactions in graphene enters as a self-energy correction for the quasi-particles. It results in an effective next-nearest neighbor hopping  $t'$ for the lattice, with strength $t'\approx 0.4$ eV. \blt{This kind of setup 
accounts for the particle-hole asymmetry,  underlying the density of states}. A magnetic field effect on the system under scrutiny has order $0.1$ T. 
Accordingly, we can also mention the lattice model associated with a spin-doped graphene layer \cite{spin}. It leads to a Haldane-like periodic intrinsic magnetic flux\footnote{\blt{Ref. \cite{marl} introduced an intrinsic local magnetic flux $\phi$, such that for a nearest-neighbor plaquette the net flux vanishes, whereas for a second nearest-neighbor plaquette there is a nonvanishing net flux. This description will be used throughout the text whenever the inter-site intrinsic magnetic flux is approached in the Haldane model.}} and also to tunable valley gaps. Although its momentum-dependent mass term has a different functional form, \blt{the second-order expansion around
Dirac points of the reciprocal lattice in the Haldane model} has the same structure as the standard Haldane model itself. Considering the investigation proposed in this work, to \textcolor{black}{properly set the discussion of our model}, a mass gap of order $\approx 0.05$ meV for graphene is considered to account for the effect of a small spin-orbit coupling of the system, experimentally established in \cite{gapi,gapi2,gapi3,gapi4,gapi5}. This effect is due to the $d$ and $p_z$ orbital mixing, which is usually neglected.  \textcolor{black}{ We will show that the existence of 
a small but non-vanishing gap, together with the influence of the hypothetical statistical interaction,  induce non-trivial topological properties for graphene in a given quasi-particle regime. As we are going to discuss, it ensures a variety of interesting physical outcomes.} \\ 
 \indent Regarding some correlated field theoretical contributions in condensed matter, there are several relevant investigations  \cite{marl,mar}. Ref. \cite{dud1} addresses the properties of graphene in the reduced\footnote{This theory is associated with a specific procedure to project the $3+1$ dimensional photon field on the planar sample instead of considering the interaction with just a purely $2+1$ dimensional field.} QED framework. Besides,  a prescription for the Coleman-like theorem for Lorentz breaking systems can be achieved. Ref. \cite{dud2} discusses the possibility of a half-quantized  Hall effect for a class of 2D gapped materials. There is also research reporting the contribution of a Chern--Simons (CS) term for chiral symmetry breaking in condensed matter systems \cite{kondo}. Recently, Ref. \cite{raya} analyzed some associated features considering a CS term embedded in the reduced QED framework. Another relevant quantum field theory investigation on the condensed matter is the description of Kekulé distortions in a graphene layer, using an extra gauge field representing the effect of a pseudo-magnetic field, also leading to a prediction of fractional charges \cite{hott}.   \\
  \indent The choice of introducing the statistical interaction is motivated by the typical low-energy (topological) setup associated with the quasi-particle phenomenology. Taking into account a generalized Hubbard--Stratonovich transformation, the fermionic self-energy interaction can be expressed with respect to a trilinear coupling with a CS field.
  The latter is well known to introduce no new degrees of freedom, with an associated vanishing Hamiltonian operator. Despite the apparent triviality of such a model, it has important implications for the overall properties of the interacting system. This is the main reason why the Kugo--Ojima--Nakanishi (KON) formalism is applied here \cite{Nak1,thooft}. Accordingly, one can also mention recent applications of this formalism as suggested in Ref. \cite{ref1}, suitably chosen to unveil subtle aspects of quantum gravity, the definition of a complementary tool for BRST symmetry extensions \cite{ref2},  the discussion of mass generation mechanisms \cite{ref3}, criteria for confinement in QCD \cite{conf}, and to unravel formal aspects of QED$_4$ in the so-called non-linear t'Hooft gauge \cite{thooft}. This formalism considers an indefinite metric quantization with a well-defined subsidiary condition to define the physical subspace. Then, it is possible to show that although the free CS model has no positive definite observable projections when it is coupled, it presents a non-trivial quantum structure, with plenty of physical implications on the quasi-particle dynamics. We also point out how the inclusion of the fermion interaction implies the emergence of positive projections for this topological field.   This quantization method is based on the Heisenberg description and has both perturbative and non-perturbative approaches  \cite{Nakpert}. These possibilities are necessary due to the quasi-gauge invariant nature of the bosonic Lagrangian. Namely, it is invariant up to adding a topological class for the case of large gauge transformations. However, the underlying physics can be invariant if a restricted class of propagator coefficients is considered. Therefore, the functional generator encoding the physical amplitudes keeps its form. This coefficient has an allowed range of values, taking our model from a non-perturbative to a perturbative structure. It motivates us to first present its general non-perturbative formulation and then consider some suitable cases to proceed with approximation schemes. This last step is associated with obtaining the self-energy, with a relevant impact on the quasi-particle dynamics. It implies renormalized solutions with topologically protected boundary states and 
 unidirectional drift velocity. Moreover, the effect of an external magnetic field on this self-interacting renormalized system implies a generalized Landau-level structure. Both solutions exhibit explicit valley asymmetry as an effect of interaction with the CS field. \textcolor{black}{ More specifically, one valley becomes topological, whereas the other one does not, leading to specific signatures in the  Landau levels and in solutions implying a spatial separation between the two valley excitations. This latter feature can define a theoretical basis for new valleytronics devices.}\\
\indent \textcolor{black}{Summing up, we accommodate our current investigation in a  proper context in the light of contemporary literature. We highlight the typical energy scale and the overall principles that constrain our efforts. Moreover, the experimental results accounting for the graphene's tiny mass define the paper's background. As we are going to address, regarding topological properties, the physics underlying a quasi-particle with a tiny mass is intrinsically different from the one related to an identically massless quasi-particle. Regarding the next sections,} the paper is organized as follows: Sec. \ref{sec2} is devoted to discussing the lattice structure of Dirac-like materials, as well as to defining the distinction from the Haldane case. Additionally, comments about the continuum field description in the low-energy limit, as well as the gauge symmetry constraints,  are provided. Sec. \ref{sec3} addresses the structure of the Poincaré group for a massless vector field in $D=2+1$ dimensions. It implies the necessity of an indefinite metric quantization. The limit of Lorentz symmetry breaking, characteristic of Dirac-like condensed matter systems, is also defined and investigated. Sec. \ref{sec4} delves into  the quantization of statistically interacting Dirac quasi-particles,  in the KON  formalism. A complete non-perturbative treatment for bosonic and fermionic fields is implemented, complying with the possibilities revealed by the gauge symmetry constraints of Sec. \ref{sec2}. In Sec. \ref{sec5}, the perturbative regime is considered, and the first-order correction for the quasi-particle self-energy is derived. Firstly, the anti-commutator structure is established. Later, in the first subsection, the Feynman version is obtained through a spectral representation. A judicious renormalization process is necessary  since the topological properties crucially depend on the relative sign between the renormalized momentum-dependent mass terms. Regarding the fermionic response associated with the self-energy, its Fourier transform contains a $\tilde p_\mu \tilde p^\mu$ momentum term with $\tilde p_\mu=(p_0, \upnu \vec p)$ denoting the specific kind of Lorentz symmetry breaking associated to these matter samples. In this case, $\upnu$ represents the quasi-particle drift velocity.  This renormalized momentum-dependent mass term generates the interaction-driven \textcolor{black}{topological nature}, leading to an effective next-nearest hopping term. Also, the renormalized structure implies that one of the valleys becomes topological, while the other does not, implying valley asymmetry. In Sec. \ref{sec6}, two relevant solutions for the renormalized system are considered. The first one is associated with topologically protected boundary states, whereas the other one contemplates the emergence of generalized Landau levels in the presence of an external magnetic field. The underlying phenomenology displays explicit valley asymmetry signatures. Finally,  conclusions and perspectives are addressed in Sec. \ref{sec7}.

\section{From the lattice to the continuum picture embedded in a topological interaction}\label{sec2}

\indent The tight-binding Hamiltonian for a honeycomb lattice model, which includes Haldane \cite{haldane} and standard Dirac-like materials, reads
\begin{align}
      H=&\sum_k C^\dagger_A(k)C_B(k)\left[-t\sum_{j=1}^3e^{i\vec k\cdot\vec d_j} \right]+\sum_k C^\dagger_A(k)C_A(k)\left[M-t_2\sum_{j=1}^3e^{i\vec k\cdot\vec e_j +i\phi}\right]  \nonumber \\ &+\sum_k C^\dagger_B(k)C_B(k)\left[-M-t_2\sum_{j=1}^3e^{i\vec k\cdot\vec e_j +i\phi}\right]     + {\rm h.\, c.},                              \end{align}
\noindent where $\phi$ represents \blt{a local intrinsic inter-site magnetic flux}, $\vec d_i$ are the vectors connecting the first nearest neighbors, namely the $A$ and $B$ sites with different crystallographic classification;  the $\vec e_i$ vectors connect the next-nearest neighbors \cite{marl} and the parameters $t$ and $t_2$  represent the nearest and next-nearest hopping couplings, respectively, whereas $C^\dagger_{A/B}(k)$ are creation operators for quasi-particles in the $A$ and $B$ sites. The $M$ parameter denotes the minimum gap between the energy bands. For standard Dirac materials, just $t$ and $M$ are non-vanishing, realizing the Wallace model \cite{wallace}. Regarding graphene, the low-energy expansion occurs around the independent valley points $k_\pm=\pm \frac{4\pi}{3a}$, presenting a quasi-particle structure. For this specific case, the Hamiltonian can be rewritten as 
\bea H=\sum_k\Psi^\dagger(k) \left(\begin{array}{ccccc}
	M& -t\displaystyle\sum_{j=1}^3e^{i\vec k\cdot\vec d_j} &       \\
	-t\displaystyle\sum_{j=1}^3e^{-i\vec k\cdot\vec d_j}  & -M&       \\
\end{array}\right)\Psi(k),  \eea
with the definition of the spinor conjugate \bea  \Psi^\dagger(k)\equiv \begin{bmatrix}
          C_A^\dagger(k),&  C_B^\dagger(k) \\
          \end{bmatrix}. \eea
          \indent Therefore, expanding the solution around the two valley points defined as $ -t\sum_{j=1}^3e^{i\vec K_{\pm}.\vec d_j} =0$, the associated Lagrangian can be derived \cite{mar,dud2}, as
                   \bea \mathcal{L}=\sum_{I=1}^2\Big(i\hbar \bar \Psi_I\gamma^\mu \tilde \p_\mu \Psi_I-m_I\bar \Psi_I \Psi_I\Big), \eea
   after a canonical transformation. The derivative appearing above is defined as $\tilde{\partial}_\mu=(\partial_0,\upnu \vec \partial)$, where $\upnu$ denotes the quasi-particle drift velocity. Here,  $\Bar \Psi\equiv \Psi^\dagger \gamma_0$, as usual, and the sum is taken over the valley degree of freedom, with $m_1=-m_2=M$.  The theory associated with the low-energy excitations assumes a Dirac-like structure in a two-dimensional faithful representation. The Dirac gamma matrices implement a matrix  representation of the Clifford algebra and can be read off as $\gamma_0=\sigma_3, \gamma_1=i\sigma_1$ and $\gamma_2=i\sigma_2$, with $\sigma_i$, for $i=1,2,3$,  denoting the Pauli matrices.\\
   \indent Considering the graphene case, the continuum field picture is valid up to the cutoff  $\frac{h{\upnu}}{a}\approx 3$ eV  with $\frac{a}{\sqrt{3}}\sim 0.142 $ nm, which is the typical length scale associated with the vector size connecting the two nearest neighbor sites in the lattice.  The quasi-particle drift velocity has order $\upnu \sim \frac{c}{300}$, with $c$ being the speed of light in the material. It depends on the microscopic lattice parameters as $\upnu\sim at$. The low-energy excitations have a tiny band gap of $|m_I|\approx 0.05$ meV \cite{gapi,gapi2}. The time-reversal invariance of the non-interacting system implies $m_1=-m_2$. Although graphene can be treated as approximately gapless, leading to interesting properties, there are also next-leading order quadratic momentum dispersion terms. \blt{Moreover, the spin-orbit coupling is responsible for the formation of a mass gap in graphene, which was experimentally measured} \cite{gapi3,gapi4,gapi5}. Since the $p_z$ orbital has zero angular momentum, this effect is due to the usually neglected $d$ and $p_z$ mixing.  \\
\indent As we are going to see, the statistical interaction generates a Haldane-like behavior driven by self-energy corrections for topologically trivial quasi-particles. The topological class associated with the phase with $|M|>3\sqrt{3}|t_2|$ and $\phi=\pi/2$, at \blt{the second-order expansion around
Dirac points of the reciprocal lattice} \cite{tese,ger}, is the one effectively reproduced by this interaction. The statistical interaction responsible for turning this model into a topologically non-trivial system reads\footnote{Here $\tilde \Box$ is defined as $\tilde{\p}^\mu\tilde{\p}_\mu$, with indices raised by the Minkowski metric.}
\bea  U=-\frac{g^2}{{\mathcal{K}}} \frac{\Big(J_1^\mu+J_2^\mu\Big)\epsilon_{\mu \gamma \nu}\tilde \partial^\gamma \Big(J_1^\nu+J_2^\nu\Big) }{\tilde \Box}.   \label{cd}   \eea
\noindent We are considering that in the free Lagrangian, as well as in the interacting one, the source of Lorentz symmetry breaking lies in the derivative/dispersion relation structure since the velocity $\upnu$ is smaller than the speed of light  $c$. The current densities appearing in Eq. (\ref{cd}) are the bilinears $J^\mu_I=\Bar \Psi_I\gamma^\mu \Psi_I $. \\
 \indent The field theory model for a self-interacting Dirac quasi-particle can be equivalently expressed through a generalized Hubbard--Stratonovich transformation\footnote{See details in  Appendix A.}
\begin{eqnarray} Z&=&N\int \prod_{\mu=0}^3\mathcal{D}{\mathcal{C}}^\mu \mathcal{D}B \prod_{I=1}^2\mathcal{D}\bar \Psi_I\, \mathcal{D} \Psi_I\times\exp\Bigg[\frac{i}{\hbar}\int d^4x\Bigg( \frac{{\cal{K}}}{2} \epsilon^{\mu \nu \beta}{\mathcal{C}}_\mu \tilde \p_\nu {\mathcal{C}}_\beta + B\tilde \partial_\mu {\mathcal{C}}^\mu+\frac{\alpha}{2}B^2 \nonumber \\&&\qquad \qquad \qquad \qquad \qquad  \qquad+\sum_{I=1}^2\Big(i\hbar\bar \Psi_I\gamma^\mu \tilde \p_\mu \Psi_I-m_I\bar \Psi_I \Psi_I+g{\mathcal{C}}_\mu \bar \Psi_I\gamma^\mu\Psi_I\Big)\Bigg)\Bigg]  \end{eqnarray}
after introducing the CS mediator, generating an analog of the internal inter-plaquette magnetic flux. It also leads to an effective nearest-neighbor hopping, through quantum corrections. 

\subsection{On the gauge symmetry discussion}

\indent  The equation of motion associated to the field $B(x)$ generates the gauge condition for the emergent local $U(1)$ symmetry. Integrating it out, the standard gauge fixing Lagrangian form is achieved.  The explicit  gauge transformations read
\bea  {\mathcal{C}}_\mu(x)\mapsto  {\mathcal{C}}_\mu(x)+\frac{1}{g}\tilde \partial_\mu \omega(x), \quad  \quad \Psi(x) \mapsto e^{i\omega(x)}\Psi(x).       \eea
\indent As it is going to be addressed, the vector field variation transforms differently than the gauge field under Lorentz transformations. However, considering only the rotational subgroup, the sum of the field and its variation transforms covariantly, defining the residual symmetry. Regarding the bosonic field, its local symmetry transformations can be 
expressed in the form
\bea {\mathcal{C}}_\mu(x) \to u^{-1}(x){\mathcal{C}}_\mu(x) u(x)+u^{-1}(x)\frac{1}{g}\tilde \partial_\mu u(x), \eea
\noindent \blt{including also the so-called large gauge transformations, consisting of the ones that do not vanish at infinity.} Up to no contributing boundary terms, the action changes as
\bea S_{CS}({\mathcal{C}}_\mu(x))\mapsto S_{CS}({\mathcal{C}}_\mu(x))+2\pi {\mathcal{K}}{\mathcal{G}}(u) \frac{\upnu^2}{g^2},\eea
with ${\mathcal{G}}(u)=\frac{1}{24\pi^2}\int d^3x \epsilon^{\mu \nu \alpha}Tr\Big({u^{-1}}\partial_\mu {u  \ u^{-1}}\partial_\nu {u\ u^{-1}}\partial_\alpha u\Big)$ denoting an integer topological invariant, which does not vanish for large gauge transformations. Although the action is not gauge invariant, the physical system can be \blt{invariant under gauge transformations. In fact, the path integral does not change by the choice }
\bea  {\mathcal{K}}=\frac{\hbar\  g^2n}{2\pi \upnu^2},     \eea
 with $n$ denoting a natural number.

\section{Indefinite metric in $D=2+1$ dimensions} \label{sec3}

\indent Analogously to what occurs in the four-dimensional case \cite{Nak1}, we prove that
 in $D=2+1$ dimensions the quantization of massless gauge vector fields also requires an indefinite metric Hilbert space.  After analyzing the Lorentzian case, the discussion on the phase\footnote{Hereon  natural units $c=\hbar=1$ are considered. Therefore, $\upnu <1$ is dimensionless.}  $\upnu <1$ associated with our specific matter content, is provided. This analysis is pertinent for this specific KON quantization scheme since the presence of the auxiliary $B$ field implies a massless or $\tilde p_\mu \tilde p^\mu$ pole\footnote{With $\tilde{p}_\mu=(p_0, v\vec p)$.}, for the free CS field in the Lorentz symmetric and Lorentz symmetry breaking frameworks, respectively.\\
\indent Considering the Lorentzian case,  the three-dimensional Poincaré group generator algebra is investigated to obtain the maximum set of compatible operators for the case of a massless particle with momentum\footnote{Implicitly assuming that the commutator operators are acting on a momentum eigenstate.} $P_\mu=(p,0,p)$, 
\bea \Big[\mathcal{J}_\mu,\mathcal{J}_\nu \Big ]=\epsilon_{\mu \nu \gamma} \mathcal{J}^\gamma,   \quad    \quad  \Big[M_{\mu \nu},P_\lambda \Big ]=i\Big(\eta_{\nu \lambda}P_\mu-\eta_{\mu \lambda}P_\nu       \Big),\quad     \quad \Big[P_{\mu },P_\lambda \Big ]=0,               \eea
in which the $SO(1,2)$ generators $\mathcal{J}^\mu=\epsilon^{\mu \nu \beta}M_{ \nu \beta}$ are defined in terms of the standard Lorentz ones.  The maximum compatible set of generators is given by $L\equiv \mathcal{J}_0+\mathcal{J}_1$ and $P_\mu$. It is worth mentioning that the orbital part  of $L$ does not contribute to this specific massless momentum frame.\\
\indent Considering the next discussions, it is useful to  consider an explicit $2+1$-dimensional representation for these generators, 
\bea  \Big(\mathcal{J}^\mu\Big)^{\alpha \beta}=\epsilon^\mu_{ \ \gamma \sigma}P^\gamma \frac{\delta }{\delta P_\sigma} \delta^{\alpha \beta}+i\epsilon^{\mu \alpha \beta}           \eea
\noindent The explicit form of the $(L)^{\alpha \beta}$ operator reads
\bea (L)^{\alpha \beta}=i\begin{pmatrix}
	0 & 1& 0     \\
	-1 & 0 &1      \\
	0& -1 & 0     \\  
\end{pmatrix}.                              \eea
\noindent These Lorentz transformations,  preserving the light-like momentum $\Lambda p=p$, are associated with $\Lambda^{\ \nu}_\mu a_\nu(\Lambda^{-1} p)=U a_\mu(p)U^\dagger$, where the $a_\mu(p)$ represents the Fourier transform of the massless vector field and $U$ denotes the  (pseudo)unitary representation of the Lorentz group.
Therefore, due to the Poincaré invariance of the vacuum state, $U|0\rangle =|0\rangle$, one concludes that the projection tensor $M_{\mu \nu} =\langle 0 |a_\mu(p) a_\nu^\dagger(p)|0\rangle$  is invariant under transformations generated by $(L)_{\alpha \beta}$ with $\Lambda^{-1} p=p$. In matrix notation, one can express
\bea M=\Lambda M  \Lambda^\intercal,      \eea
for $\Lambda^\intercal$ denoting 
 the transpose of $\Lambda$.\\
\indent Considering the infinitesimal form of the Lorentz transformation  $\Lambda_{\mu \nu}=\delta_{\mu \nu}+i(L)_{\mu \nu}\varepsilon$, with $\varepsilon \to 0$, yields $LM=ML$. This condition implies the following general form
\bea M_{\mu \nu}=\begin{pmatrix}
    	a& 0& c     \\
	 0& (a-c) & 0    \\
	c& 0 & a  \end{pmatrix}+\begin{pmatrix}
0 & b& 0     \\
-b & 0 &b     \\
0& -b & 0    
\end{pmatrix}                  \eea
for the projection tensor. Due to the Hermitian nature of the projection matrix, the elements from the symmetric part must be real, and the ones from the antisymmetric sector must be purely imaginary. The emergence of this antisymmetric part is associated with the possibility of adding a CS term in $2+1$ dimensions.  The eigenvalues of the projection matrix can be expressed as  $\lambda=a+c $ and $\lambda=a-c \pm |b|$,  with $|b|$ being the modulus of the complex number $b$. This result means that the eigenvalues of the matrix projections are indeed indefinite. \\
\indent Regarding the phase $\upnu<1$, the very reason for the Lorentz breaking in these quasi-particle systems is the fact that the Hamiltonian presents the generalized momentum/derivative structure 
\bea \tilde \partial_\nu \mapsto \tilde{\Lambda}_\mu^{\ \nu}\tilde \partial_\nu, \eea
\noindent transforming covariantly under $\tilde{\Lambda}_\mu^{\ \nu}$ and not under the Lorentz transformations,  $\Lambda_\mu^{\ \nu}$, while the remaining vectors present in this Hamiltonian are tensor representations of the Lorentz group. Since the underlying symmetry of nature is Lorentzian, its action on the components of  $\tilde \partial_\mu$ defines   $\tilde{\Lambda}_\mu^{\ \nu}$ as
\bea   \tilde{\Lambda}_\mu^{\ \nu}= \Omega_{\mu}^{\ \alpha} \Lambda_\alpha^{\ \gamma}(\Omega^{-1})_\gamma^{\ \nu}          \eea
with \bea  \Omega_{\alpha}^{\ \beta}=\left(\begin{array}{ccccccc}
	1& 0& 0     \\
	 0& v & 0     \\
	0& 0 & v     \\  
	\end{array}\right)\eea
\indent Regarding the Lorentz symmetry breaking case, with $\upnu\leq 1$, the reduced symmetry is associated to rotations\footnote{For them, $\tilde \Lambda=\Lambda$. } and translations. The generator algebra becomes
\bea  \Big[M_{12},P_\lambda \Big ]=i\Big(\eta_{2 \lambda}P_1-\eta_{1 \lambda}P_2      \Big),\quad     \quad \Big[P_{\mu },P_\lambda \Big ]=0.               \eea
The spatial rotations on the plane are associated with the generator $(J^0)^{\alpha \beta}$. The compatible set of generators is just the set of $P_\mu$ components. They have no direct implication in the form of the projection matrix. However, nothing avoids using the indefinite metric quantization as an organizing principle to investigate this kind of system.

\section{The $B$ field quantization} \label{sec4}

\indent In this section, the indefinite metric quantization in the KON framework is developed to investigate some non-perturbative aspects of the statistical interaction for quasi-particles. The Lagrangian reads
\begin{equation}
\mathcal{L} = \frac{{\cal{K}}}{2} \epsilon^{\mu \nu \beta}{\mathcal{C}}_\mu \tilde \p_\nu {\mathcal{C}}_\beta + B\tilde \partial_\mu {\mathcal{C}}^\mu+\frac{\alpha}{2}B^2 +\sum_{I=1}^2\Big(i\bar \Psi_I\gamma^\mu \tilde \p_\mu \Psi_I-m_I\bar \Psi_I \Psi_I+g{\mathcal{C}}_\mu \bar \Psi_I\gamma^\mu\Psi_I\Big),
\label{AlphaGauge}
\end{equation}
with ${\cal{K}}=\frac{g^2n}{2\pi\upnu^2}$. The sum is taken over the valley's degree of freedom.  The term proportional to $\alpha$ is associated with a generalization of the Lorentz gauge.\\
\indent The operator equations of motion read
\begin{eqnarray}
     {\cal{K}}\epsilon_{\mu \nu \rho}\tilde \p^\nu {\mathcal{C}}^{\rho}(x)&=&-g\sum_{I=1}^2\bar \Psi_I(x)\gamma_\mu\Psi_I(x)+\tilde \partial_\mu B(x),\\
     \tilde \p_\mu {\mathcal{C}}^\mu(x)&=&-\alpha B(x), \\
 \Big[i\gamma^\mu\Big(\tilde \p_\mu +ig{\mathcal{C}}_\mu(x)\Big)-m_I\Big]\Psi_I(x)&=&0,\\ \bar \Psi_I(x)\Big[i\gamma^\mu\Big(  \overset{\leftarrow}{\tilde \partial_\mu}               -ig{\mathcal{C}}_\mu(x)\Big)+m_I\Big]&=&0, \end{eqnarray}
being compatible with the quasi-particle current density conservation
\bea \tilde \p_\mu\Big(\bar \Psi_I(x)\gamma^\mu\Psi_I(x)\Big)=0.\eea
It implies that the effective current operator is indeed $J_i^{eff}(x)=\upnu \bar \Psi_I(x)\gamma_i\Psi_I(x) $ and the charge density reads $\rho=\bar \Psi_I(x)\gamma_0\Psi_I(x)$.\\
\indent Taking the divergence of the photon field equation of motion yields
\bea \tilde\Box B(x)=0.\eea
\noindent The positive semi-definite metric Hilbert subspace ${\cal{V}}_{phys}$ is defined in terms of the positive frequency part of the auxiliary $B(x) $ field as \cite{Nakpert}
\bea B^+(x)|\,phys\rangle=0, \quad \quad \forall\; |phys\rangle \ \in {\cal{V}}_{phys},\eea
being a Poincaré invariant definition. Since zero-norm states have always null projections in ${\cal{V}}_{phys}$, the observable physical subspace can be properly defined as the quotient space
\bea {\cal{H}}_{phys}=\frac{\Bar{{\cal{V}}}_{phys}}{{\cal{V}}_0}, \eea  
with $\mathcal{V}_0$ representing the zero norm subspace.\\
\indent To derive the quantum field  (anti)commutators, one starts by calculating their initial conditions by the correspondence principle. To this end, the knowledge of the canonical momenta           \begin{align}
\pi^i(x) &= \frac{\delta{\cal{L}}}{\delta(\p_0{\mathcal{C}}_i(x))} = \frac{{\cal{K}}}{2}\epsilon^{ij}{\mathcal{C}}_j,\quad \quad \pi^0(x) = \frac{\delta{\cal{L}}}{\delta(\p_0{\mathcal{C}}_0(x))} = B(x), \quad \pi_B(x) = \frac{\delta{\cal{L}}}{\delta(\p_0 B(x))} = 0\\
\pi_{\Psi_I}(x) &= \frac{\delta{\cal{L}}}{\delta(\p_0 \Psi_I(x))} = i\bar \Psi_I(x)\gamma^0 , \quad \pi_{\bar \Psi_I}(x)=\frac{\delta{\cal{L}}}{\delta(\p_0 \bar \Psi(x))}=0
\end{align}
has fundamental importance.
 The presence of the auxiliary $B(x)$ field turns the system into a second-class one. It implies that the only constraints are the primary ones. They can be considered in strong form, avoiding any kind of ambiguity if the Dirac brackets are considered. Therefore, this procedure leads to a well-defined reduced phase space. Then, the correspondence principle can be properly established.\\
\indent The  Berezin brackets have the following structure\footnote{Since the system is of second class from the beginning, the unusual factor $\frac{1}{2}$ appears when one considers the Dirac brackets, see Appendix $B$. Here, we do not consider the subscript $D$ for simplicity.}
\begin{align}
\big\{{\mathcal{C}}_i(x),\pi^j(y)\big\} &= \frac{1}{2} \delta_i^j \delta^2(x-y) ,\quad \big\{{\mathcal{C}}_0(x),B(y)\big\} = \delta^2(x-y), \quad  
\big\{\Psi_I(x),i\bar \Psi_I(y)\gamma^0\big\} = \delta^2(x-y). 
\end{align}
\noindent Due to the correspondence principle and  the operator equations of motion, the non-vanishing initial conditions for the (anti-)commutators are obtained,
\begin{subequations}
\begin{align}
\Big\{  \Psi_I(x), \bar\Psi_I(y)\Big \}_0 &= \gamma^0\delta^2(x-y),\quad \qquad \ \  \Big[{\mathcal{C}}_i(x),{\mathcal{C}}_j(y)\Big ]_0 = \frac{i}{{\cal{K}}}\epsilon_{ij}\delta^2(x-y),  \\ 
\Big[{\mathcal{C}}_0(x), B(y)\Big]_0 &= i\delta^3(x-y),\qquad \quad \
\Big[{\mathcal{C}}_0(x), \p_0{\mathcal{C}}_0(y)\Big]_0 = -i \alpha \delta^2(x-y), \\
\Big[B(x),\p_0B(y)\Big]_0 &= 0,\quad \quad \qquad \quad \qquad \ 
\Big[{\mathcal{C}}_\mu(x),\p_0B(y)\Big]_0 = -i\tilde \p_k^y\delta_\mu^k\delta^2(x-y), \\
\Big[\p_0{\mathcal{C}}_i(x),{\mathcal{C}}_0\Big]_0&=i\frac{\epsilon_{ij}}{{\cal{K}}}\tilde \p^j \delta^2(x-y), \quad \quad \ \
\Big[\p_0{\mathcal{C}}_0(x),{\mathcal{C}}_i\Big]_0=-i\frac{\epsilon_{ji}}{{\cal{K}}}\tilde \p^j \delta^2(x-y),\\
\Big[ \Psi_I(x), \p_0B(y)\Big]_0&=e\Psi_I(x)\delta^2(x-y), \quad
\Big[ \bar \Psi_I(x), \p_0B(y)\Big]_0=-e\bar \Psi_I(x)\delta^2(x-y),
\end{align}
\end{subequations}
for which the subscript $0$ denotes quantities evaluated at equal times.\\
\indent Before proceeding with the analysis, it is worth mentioning that for every operator $F(x,y)$ obeying $\hat{O}^x F(x,y)=G(x,y)$, with $\hat {O}^x$ being a given second order differential operator acting in coordinate $x$, there is the following integral representation \cite{Nakpert}
\begin{align}
F(x,y)&= \int d^3u \ \varepsilon(x,y,u) \tau(x-u)G(u,y) \nonumber \\  
\quad&\;\;\;\;\;-\int d^2u \Big[ \tau(x-u)\p_0^uF(u,y)-\p_0^u\tau(x-u)F(u,y) \Big]_{u^0=y^0},
\label{ghostTwoPoint}
\end{align}
with $\tau(x-y)$ being an operator valued distribution such that $\hat{O}^x\tau(x-y)=0$. The symbol $\varepsilon(x,y,u)$ is defined in terms of Heaviside functions as $ \Theta(x_0-u_0)- \Theta(y_0-u_0)$.  The integral representation for the  case of a first-order differential operator ${\cal {O}}^x$ associated with the Dirac equation with ${\cal {O}}^x f(x,y)=g(x,y)$, reads
\begin{align}
f(x,y)&= \int d^3u \ \varepsilon(x,y,u) S(x-u)g(u,y)  
+i\int d^2u \Big[ S(x-u)\gamma^0f(u,y) \Big]_{u^0=y^0},
\end{align}
with $S(x-u)$, which is going to be further defined, obeying ${\cal {O}}^x S(x-u)=0$. Therefore the $B$-field commutators can be evaluated by employing the integral representation
\begin{align} \Big[B(x),B(y)       \Big]=&0 , \qquad \qquad \qquad \ \quad \quad \Big[{\mathcal{C}}_\mu(x),B(y)       \Big]=i\tilde \p_\mu D(x-y,0), \\
\Big[\Psi_I(x),B(y)       \Big]=&e\Psi_I(x)D(x-y,0) ,\qquad  \Big[\bar \Psi_I(x),B(y)       \Big]=-e\bar \Psi_I(x)D(x-y,0),
\end{align}
with the distribution appearing above $D(x-y,0) $ being the generalization of the so-called massless Pauli--Jordan one defined in Appendix C. Although its Fourier transform reveals a Lorentz-violating pattern, it can still be decomposed into positive/negative frequency parts. The dependence on the difference of the coordinates is due to the $2+1$ dimensional translational symmetry of the system's Hamiltonian.\\
\indent The free limit is the first one to be analyzed in this paper. Contracting the operator $\epsilon_{\mu \nu \beta}\tilde \p^\beta$ with the vector boson field equation defining the generalized gauge curvature leads to
\bea \tilde \Box_x \tilde \Box_y \Big[ {\mathcal{C}}_\mu(x),{\mathcal{C}}_\nu(y) \Big]=0\eea
considering the zero norm character of the $B$ field. The initial conditions  and the equations of motion yield
\bea \Big[ {\mathcal{C}}_{\rho}(x),{\mathcal{C}}_{\beta}(y)\Big]=-\frac{i}{{\cal{K}}}\epsilon_{\rho  \beta \mu}\tilde \p^\mu D(x-y,0)-i\alpha \tilde \p_\rho \tilde \p_\beta E(x-y, 0),  \eea
considering the integral representation. The gauge field components have null norm and non-positive definite projections between them. The new distribution  $E(x-y,s)$ introduced above, associated with double pole equations, is also defined in Appendix C. \\ 
\indent For the free Dirac field case, considering the integral representation, one obtains the anti-commutator\footnote{For every spacetime vector, $A_\mu\gamma^\mu$ is denoted as $\slashed{A}$.}
\bea (i  \tilde {\slashed{\p}}-m_I)_x\Big\{\Psi_I(x),\bar \Psi_I(y)                    \Big\}=0\eea
\noindent Hence, taking into account the initial conditions, the integral representation furnishes
\bea \Big\{\Psi_I(x),\bar \Psi_I(y)\Big\}=i(i\tilde{\slashed{\p}}+m_I)D(x-y,m^2_I)\equiv iS_I(x-y).\eea
\noindent A generalized massless pole field obeying the subsidiary condition is given by ${\cal{A}}_\mu(x)={\cal{K}}\epsilon_{\mu \nu \alpha}\tilde \p^\nu {\mathcal{C}}^\alpha(x)$. Its norm, in the free case, vanishes
\bea \langle 0 |\Big[{\cal{A}}^\mu(x),{\cal{A}}^\nu(y)\Big]|0\rangle=0.\eea
\indent The pole field ${\cal{A}}^\mu(x)$ is gauge invariant, since it commutes with $B(x)$, being a physical field. However, it occupies the null norm sector of the theory. This complies with the fact that this model has no local degrees of freedom. However, when the interaction is turned on, this observable field enters the positive definite subspace of the Hilbert space.  This is closely related to the current-current commutator structure. Then, the vanishing of the commutator 
\bea \Big[\bar \Psi_I(x)\gamma^\mu \Psi_I(x),B(y)          \Big]=0\eea
implies that the current is an observable, as it should be. This object indeed has positive projections. \\
\indent The interacting gauge invariant field ${\cal{A}}^\mu(x)$ is associated to the commutator
\bea \langle 0 |\Big[{\cal{A}}^\mu(x),{\cal{A}}^\nu(y)\Big]|0\rangle=g^2\langle 0 |\Big[ J^\mu(x),J^\nu(y)\Big]0\rangle, \eea
with $ J_\mu(x)\equiv \sum_I\bar \Psi_I(x)\gamma_\mu \Psi_I(x)$.  This result shows the emergence of a vector field positive norm due to the presence of the interaction. However, the observable field commutator is algebraically related to the one associated with the sources. Therefore, there is no independent asymptotic boson field, in accordance with the topological nature of this auxiliary field.\\
\indent Regarding the commutator between the vector fields, it is possible to show that the equations of motion imply that
\bea \tilde \Box_x \tilde \Box_x \Big[{\mathcal{C}}_\mu(x), 
{\mathcal{C}}_\nu(y)\Big]=\frac{g^2}{{\cal{K}}^2}\epsilon_{\mu \sigma \beta}\epsilon_{\nu \omega \gamma}\tilde \p^\sigma_x \tilde \p^\omega_x\Big[J^\beta(x), J^\gamma(y)\Big]. \eea
\noindent Therefore, using the integral representation formula twice, and considering the initial conditions,  the general solution can be derived, as
\begin{align}
\langle 0 |\Big[{\mathcal{C}}_\mu(x&), {\mathcal{C}}_\nu(y)\Big]|0\rangle =-\frac{i}{{\cal{K}}}\epsilon_{\mu  \nu \alpha}\tilde\p^\alpha D(x-y,0)-i\alpha \tilde \p_\mu \tilde \p_\nu E(x-y,0)  \nonumber\\-&\frac{g^2}{{\cal{K}}^2}\epsilon_{\mu \sigma \beta}\epsilon_{\nu \omega \gamma}\tilde \p^\sigma_x \tilde \p^\omega_x\int d^3\omega d^3u \ \epsilon(y,x,
u) \epsilon(x,u,\omega) D(x-\omega,0)D(y - u,0)\langle 0 |\Big[J^\beta(\omega),J^\gamma(u)\Big]|0\rangle 
\end{align}
in which the vacuum expectation value of the current commutator defines the causal version of the boson self-energy, see Appendix C.\\
\indent The equations of motion for the quasi-particle fields yield
\bea\Big(i\tilde{\slashed{\p}}^x -m_I\Big) \Big\{\Psi_I(x),\bar \Psi_I(y) \Big\} \Big(i  \overset{\leftarrow}{\tilde{\slashed{\p}}}^y               +m_I\Big)=-g^2\Big\{\gamma^\mu {\mathcal{C}}_\mu(x)\Psi_I(x), \bar \Psi_I(y){\mathcal{C}}_\nu(y)\gamma^\nu  \Big\} \label{33} \eea
with the left-hand side denoting the quasi-particle self-energy in its anti-commutator version. The complete solution is then obtained through the initial data and the integral representation for the anti-commutator (\ref{33}). It explicitly reads
\begin{align}\Big\{\Psi_I(x),\bar \Psi_I(y) \Big\}=&iS_I(x-y)-\int d^3\omega d^3u \ \epsilon(y,x,u) \epsilon(x,u,\omega) S_I(x-\omega){\Sigma_I}(\omega,u)S_I(u-y)\nonumber \\ 
&-ig\int d^3\omega \epsilon(y,x,\omega) S_I(x-\omega) \slashed{\mathcal{C}}(\omega)S_I(w-y). \end{align}
\noindent Since this bosonic field is just an artifice to linearize the fourth-order fermion interaction, it is considered to vanish on average.    Therefore, the vacuum expectation value reads
\bea\langle 0 |\Big\{\Psi_I(x),\bar \Psi_I(y) \Big\}|0\rangle=iS_I(x-y)\!-\!\int d^3\omega d^3u \epsilon(y,x,u) \epsilon(x,u,\omega) S_I(x\!-\!\omega)\Sigma_I(\omega\!-\!u)S_I(u\!-\!y), \eea
with
$\Sigma_I(x-y)\equiv g^2\langle 0 |\Big\{\gamma^\mu {\mathcal{C}}_\mu(x)\Psi_I(x),\bar \Psi_I(y){\mathcal{C}}_\nu(y)\gamma^\nu  \Big\}|0\rangle$.\\
\indent There is no correlation between different valleys. To prove it the  Gell-Mann and Low theorem can be applied to relate the complete  propagator to the expression on the interaction picture, associated  with free fields, denoted by the superscript $0$ in the interaction picture vacuum $|\Omega \rangle$, as
\begin{equation}
    \langle0|T\Psi_1(x)\bar \Psi_2(y)|0\rangle =\frac{\langle \Omega|T \Psi_1^0(x)\bar \Psi_2^0(y)\exp{\Big\{i\sum_Ig\int d^4x (\Bar{\Psi}_I^0\gamma^\mu \Psi_I^0 {\mathcal{C}}_\mu^0)\Big\}}|\Omega\rangle}{\langle \Omega|T\exp{\Big\{i\sum_Ig\int d^4x (\Bar{\Psi}_I^0\gamma^\mu \Psi_I^0 {\mathcal{C}}_\mu^0)\Big\}}|\Omega \rangle}.
\end{equation}
Considering the Wick theorem and the fact that there are no free Feynman correlators between the $\Psi_1(x)$ and $\Psi_2(x)$ operators, it is possible to conclude that the complete propagator vanishes for all orders,  implying valley independence.  The condition regarding the free Feynman propagator comes from the fact that  the free $\Psi_1(x)$ and its dual anti-commute with both $\Psi_2(x)$ and its correlated free dual operator. Since the free Feynman function can be expressed by an integral representation in terms of the anti-commutator distribution, it also vanishes. This theoretical prediction is a relevant property to be taken into account in the development of valleytronics devices.

\section{ On the quasi-particle self-energy }\label{sec5}

\indent The non-perturbative expression for the anti-commutator version of the quasi-particle self-energy reads
\begin{align}     \Sigma_I(x-y)=g^2\langle 0 |\Big\{\gamma^\mu {\mathcal{C}}_\mu(x) & \Psi_I(x), \bar \Psi_I(y){\mathcal{C}}_\nu(y)\gamma^\nu   \Big\}|0\rangle. \end{align} 
\noindent The first approximation for the self-energy is obtained through the free (anti)commutators 
\begin{align} \Sigma_I(x-y)=-\frac{g^2}{{\cal{K}}}\gamma^\mu       \Big(\epsilon_{\mu  \nu \alpha}\tilde \p^\alpha D^+(x-y,0)S^+_I(x-&y)   -\epsilon_{\mu  \nu \alpha}\tilde \p^\alpha D^-(x-y,0)S^-_I(x-y)  \Big) \gamma^\nu,                     \end{align}
where, for simplicity, just the gauge-independent parts are being highlighted\footnote{These are the ones that contribute to the physical amplitudes.}. This function is a superposition of positive and negative frequency parts. \\
\indent In order to consider perturbation theory, the CS  coefficient must be in the range $n\geq 10$. Although the strength of the coupling $g$ is  dependent on the specific interaction generating mechanism, a perturbative regime necessarily demands $g<\upnu$. This is a condition to ensure a perturbative nature for both bosonic and fermionic self-energies, see Appendix C. In this setup, $n\approx 100$ is considered. Although it seems to be an \textit{ad-hoc} fixation to ensure a good convergence, this is, in fact, a conservative approach. As it is going to be demonstrated, this is the minimum value to ensure radiative corrections that are small compared to the free contributions according to a naturalness principle. Moreover, Ref. \cite{tese} demonstrates that the coefficient of the momentum square term characterizing \blt{the second-order expansion around
Dirac points of the reciprocal lattice in the Haldane model} is given by $a^23\sqrt{3}t_2$, with $a$ associated with the lattice spacing and $t_2$ being the second-nearest neighbor hopping parameter. For the present case of graphene, considering an induced hypothetical hopping parameter of order $t_2\approx 0.1$ eV, as commonly assumed for Haldane-like lattices \cite{marl,haldane}, the mentioned coefficient has order $10^{-6}$  eV$^{-1}$. The radiative corrections due to the present model can effectively recover this typical magnitude just for $n\approx 10^4$. However, our intention here is just to reproduce the qualitative topological properties of a Haldanized model at the low energy approximation.  \\
\indent The positive frequency part associated with the anti-commutator version of the self-energy function can be expressed in momentum space as
\begin{multline}\qquad \qquad \quad \Sigma^+_I(\tilde p)=\frac{  ig^2}{(2\pi)^3 {\cal{K}}}\int\ d^3q \gamma^\mu \epsilon_{\mu \nu \rho}(\tilde p^\rho-\tilde q^\rho)D^+(\tilde p-\tilde q,0)(\slashed{\tilde q}+m_I)D^+(\tilde q,m^2_I)\gamma^{\nu},             \end{multline}
with $ d^3q=\frac{1}{\upnu^2}d^3\tilde q$ and $D^+(\tilde q,m^2_I)$ being the Fourier transform of the generalized Pauli-distribution explicitly displayed in Appendix C.\\
\indent Considering the Dirac gamma matrices identities in Appendix D, this expression can be evaluated as
\begin{align}\Sigma_I(\tilde p)=\frac{g^2}{4{\cal{K}}\upnu^2 \sqrt{\tilde p^2}}\left[ -(\slashed{\tilde p}+m_I)\left(1+\frac{m^2_I}{\tilde p^2}\right)\slashed{\tilde p}+2m^2_I+2\slashed{\tilde p}m_I   \right]\theta(\tilde p^2-m^2_I)\sign(p_0),\end{align}
with $\Sigma_I(\tilde p)=\Sigma_I^+(\tilde p)+\Sigma_I^-(\tilde p)$.\\
\indent As mentioned in Sec. \ref{intro11}, the physical properties of the solution strongly depend on the relative sign associated with the radiatively generated momentum squared terms. Therefore, to obtain its Feynman version, useful to describe several physical processes, a judicious analysis in terms of well-defined spectral representation is necessary. \textcolor{black}{It is worth mentioning that this distribution can be associated with the imaginary part of the Feynman self-energy. It defines the system's spectral function \cite{spec} width and controls the quasi-particle lifetime. In this specific case, this imaginary part vanishes on-shell, implying a long-lived excitation. The spectral function, as well as some self-energy signatures associated with interacting quasi-particles, can be experimentally revealed by techniques such as ARPES \cite{arpes}, see the recent achievements. We also highlight the fact that although the mass term is small and often ignored in some well-established approximation schemes, regarding topology, even a tiny parameter can define different physical phases. }\\

\subsection{On the Feynman self-energy}

\indent In what follows, obtaining the Feynman version of the self-energy is mandatory. Several physical observables are built utilizing this radiative correction. More specifically, we are interested in the renormalized fermionic two-point effective action. For this purpose, the spectral representation is a useful tool. The associated spectral density is related to the anti-commutator version of the self-energy as
\bea \Sigma_I(\tilde p)=(2\pi) \rho_I(s=\tilde p^2,\tilde{\slashed{p}} )\sign(p_0).  \eea
\indent This relation can be straightforwardly obtained from the integral representation 
\bea     \Sigma_I(x-y)=i\int ds\ \rho_I(s,\slashed{\tilde p})D(x-y,s). \eea
\noindent Then, replacing the Pauli--Jordan function $iD(x-y,s)$ by $D_F(x-y,s)$, with the latter denoting the Feynman distribution, the time-ordered version of the self-energy is readily derived. The subtracted version of the spectral representation for the Feynman self-energy reads
\begin{multline} -i\Sigma^F_I(\tilde p)= -\int ds \frac{i\tilde p^2}{s(s-\tilde p^2-i0)}\frac{g^2}{8\pi{\cal{K}}\upnu^2 \sqrt{s}}\left[ -\left(1+\frac{m^2_I}{s}\right)(s+m_I\slashed{\tilde p})+2m^2_I+2\slashed{\tilde p}m_I   \right]\theta(s-m^2_I),                   \end{multline}
with subtraction point $\tilde p_\mu=0$, where $\Sigma^F_I(0)=0$. This procedure is necessary to define a convergent representation. This is justified by the fact that the Lagrangian parameters can indeed be tuned to achieve this structure. In fact, this is just an intermediate step to define a useful alternative integral representation. In order to achieve the on-shell renormalization conditions, one can perform a finite renormalization on the subtracted structure.\\
\indent It is possible to prove that this subtracted object is equivalent to the useful formula 
\bea i\Sigma_I^F(\tilde p)=\frac{i}{2\pi}\int_{-\infty}^{\infty}dt\frac{\Sigma_I(t\tilde p)}{t^2(1-t)},\eea
which can be evaluated in terms of simple partial fraction identities. Hence, the first approximation for the renormalized Feynman self-energy can be derived from the latter functional with renormalized parameters and by the addition of two extra terms arising from the bare action
\begin{eqnarray}  i\Big[\Sigma_I^F(\tilde p)\Big]^R&=&(m^R_I)^2\frac{g^2}{4{\cal{K}}\upnu^2\sqrt{\tilde p^2}}\frac{i}{2\pi}\Bigg[\log\left({\frac{1-\sqrt{\frac{\tilde p^2}{(m^R_I)^2}}}{1+\sqrt{\frac{\tilde p^2}{(m^R_I)^2}}}}\right)+2\sqrt{\frac{\tilde p^2}{(m^R_I)^2}}     
 \Bigg]\nonumber\\&&+\slashed{\tilde p}m_I^R\frac{g^2}{4{\cal{K}}\sqrt{\tilde p^2}\upnu^2}\frac{i}{2\pi}\Bigg[ \log\left({\frac{1-\sqrt{\frac{\tilde p^2}{(m^R_I)^2}}}{1+\sqrt{\frac{\tilde p^2}{(m^R_I)^2}}}}\right)+2\sqrt{\frac{\tilde p^2}{(m^R_I)^2}}           \Bigg]\nonumber\\&& -\frac{\slashed{\tilde p} (m^R_I)^3g^2}{4{\cal{K}}\upnu^2\sqrt{\tilde p^2}p^2}\frac{i}{2\pi}\Bigg[ \log\left({\frac{1-\sqrt{\frac{\tilde p^2}{(m^R_I)^2}}}{1+\sqrt{\frac{\tilde p^2}{(m^R_I)^2}}}}\right)+2\sqrt{\frac{\tilde p^2}{(m^R_I)^2}}        +\frac{2}{3}\left(\frac{\tilde p^2}{(m^R_I)^2}\right)^{\frac{3}{2}}  \Bigg] \nonumber\\&&-\frac{\tilde p^2g^2}{4{\cal{K}}\upnu^2\sqrt{\tilde p^2}}\frac{i}{2\pi }\log\left({\frac{1-\sqrt{\frac{\tilde p^2}{(m^R_I)^2}}}{1+\sqrt{\frac{\tilde p^2}{(m^R_I)^2}}}}\right)   +C\slashed{\tilde p}+C',              \end{eqnarray}
with counterterms that follow from a finite renormalization of the action, where $C=\delta_2$ and $C'=-(\delta_2+\delta_m)m_I^R$, since the fermion field is renormalized as $\Psi_I(x)=\sqrt{Z_2}\Psi^R_I(x)$ and the mass parameter as $m_I=Z_m m_I^R$ with $Z_2=1+\delta_2$ and $Z_m=1+\delta_m$. \textcolor{black}{ Therefore, one can constrain the renormalization factors to yield}
\begin{align} i\Big[\Sigma_I^F(\tilde p)\Big]^R=\frac{ig^2}{4\pi {\cal{K}}\upnu^2 }\Big(\slashed{\tilde p}-m^R_I\Big)\frac{ \slashed{\tilde p}}{m^R_I}\mathcal{F}(\tilde{p}^2)   +i\tilde C(\slashed{\tilde p}-m^R_I),                                \end{align}
\noindent \textcolor{black}{with the definition}
\bea \mathcal{F}(\tilde{p}^2)= \left[\frac{(m^R_I)^2\sign(m^R_I)}{\tilde p^2}+\frac{\left(m^R_I\right)^3}{2\tilde p^2\sqrt{\tilde p^2}}\left(1-\frac{\tilde p^2}{(m^R_I)^2}\right)\log\left(\frac{1-\sqrt{\tilde p^2/(m^R_I)^2}}{1+\sqrt{\tilde p^2/(m^R_I)^2}}\right)      \right], \eea
\noindent \textcolor{black}{ensuring the pole at the renormalized mass by the condition}
\bea    \Big[\Sigma_I^F(\tilde p)\Big]^R\Big \vert_{\tilde p=m_I^R}=0.\eea
\indent \textcolor{black}{Although planar fermionic self-energies frequently display the so-called on-shell singularities \cite{shell}, taking a careful limiting procedure, one can show that the potentially problematic logarithmic function does not contribute to this specific limiting procedure. Different from renormalized fermions interacting through a standard bosonic mediator \cite{pim1,pim2}, the statistical interaction leads to a well-defined on-shell limit}.\\
\indent \textcolor{black}{As it will be addressed, this renormalized structure has another peculiarity. Therefore, in order to define a consistent truncated regime, we present some preliminary considerations. According to Ref. \cite{shen}, this peculiar aspect is associated with an indication of a possible emerging topological phase in the quasi-particle regime. It is related to the sign change in the momentum-dependent renormalized mass term when evaluated at zero and at the infinite spatial momentum in a zero frequency surface, see the next discussions on the so-called topological Hamiltonian. Then, valleys with this change in gap sign are distinct from the ones in which it remains the same}\footnote{Here, the operator  $\mathcal{R}$ takes the real part of a function.}
\bea    \mathcal{M}(\vec p)=(1-\tilde C)m_I^R+\mathcal{R}\Bigg\{\frac{ \vec{p}^{\,2}\,g^2}{8\pi{\cal{K}}\upnu \sqrt{-\vec p^2}}\log\left({\frac{1-\upnu \sqrt{\frac{- \vec p^2}{(m^R_I)^2}}}{1+\upnu\sqrt{\frac{- \vec p^2}{(m^R_I)^2}}}}\right)\Bigg\}.      \eea
\noindent Therefore, for the limit $|\vec p|\to 0$ one gets 
\bea  \mathcal{M}(0)=(1-\tilde C)m_I^R,\eea
while at $|\vec p| \to \infty$ one has\footnote{Rewriting the logarithm as $\log\Big({\sqrt{\frac{(m^R_I)^2}{-\upnu^2|\vec p|^2}}-1}\Big)-\log\Big({\sqrt{\frac{(m^R_I)^2}{-\upnu^2|\vec p|^2}}+1}\Big)$ and using $\lim_{\epsilon \to 0}\log\Big({-|A|-i\epsilon}\Big)=\log{|A|}-i\pi $. Considering our quasi-particle setup and the associated orders of magnitude, the infinity considered here is to be properly understood as a configuration approaching the cutoff scale discussed in the second section.}
\bea  \mathcal{M}(|\vec p|\to \infty)\approx -\lim_{|\vec p|\to \infty} \frac{g^2\sqrt{|\vec p|^2} }{8\mathcal{K}\upnu }.\eea
 \noindent   \textcolor{black}{Hence, if }
\bea \frac{(1-\tilde C)m_I^R}{\frac{g^2 }{8\mathcal{K}\upnu }}>0,\label{ratio}\eea  \textcolor{black}{a topological phase can, at principle, be possibly reached. Consequently, in this case, there is a  critical momentum configuration that closes the momentum-dependent mass gap. Interestingly, one can perform a discretization for this continuous approximation \cite{shen,discret}. The linear and quadratic momentum terms are replaced by proper periodic functionals involving them and the lattice spacing. In this case, one would conclude that the map relating the Brillouin zone torus to the two-sphere, associated with our model, possibly implies a non-trivial topology depending on the magnitude of the system's parameters. It is another indication that this model may have a special status and can furnish a ground for a well-defined approachable truncation outlined in the next paragraphs.  \\
\indent Considering these facts as preliminary motivations, we will focus on a specific regime near the mass shell to explicitly evaluate if it has indeed a topological nature.
Finally, it is worth mentioning that the whole momentum-dependent mass term vanishes for an identically gapless quasi-particle. Then, as previously mentioned, considering the experimental results \cite{gapi,gapi2,gapi3,gapi4,gapi5}, graphene has indeed a tiny gap that should be accounted for 
 correctly describing the topological properties of the system. There is also a set of situations for which the massless approximation works pretty well characterizing several properties of this system as, for example, the restriction of backscattering for its electron constituents \cite{marl}. However, when regarding topological properties, a small gap and an identically vanishing one lead to completely different properties.}   \\
\indent  \textcolor{black}{In order to fix the unitary residue for the pole}
\bea \frac{\partial \Big[ \Sigma_I^F(\tilde p)\Big]^R}{\partial \slashed{\tilde p}}\Bigg \vert_{\tilde p=m_I^R}=0,\eea
\textcolor{black}{one should have $\tilde C=-\frac{g^2}{4\pi \mathcal{K} \upnu^2}{\rm sign}(m^R_I)$. Therefore, since the imaginary part vanishes on-shell, these on-shell excitations appear as a delta-like distribution in the corresponding spectral function, being long-lived.} \\
    \indent \textcolor{black}{From now on, a simple polynomial approximation capable of retaining the main properties of the system is developed. We derive a scheme in which just small deviations in $\tilde p_\mu \tilde p^\mu$ from the on-shell configuration are considered. Therefore, the imaginary part of the self-energy in this regime is non-vanishing but still small, defining configurations with a considerable contribution to the spectral density. Moreover, by the proper structure of the self-energy, one easily concludes that it is at least of the first order in this deviation. The criterion is the following. Considering deviations much smaller than the mass scale, the tiniest contributions to be considered are the quadratic ones divided by just one power of the renormalized mass. Contributions of higher powers are ignored. Therefore, for $\mathcal{F}(m_I^R)\approx {\rm sign}(m_I^R)$, retaining just the terms that would lead to this class of deviations, the self-energy becomes }
\bea i\Big[\Sigma_I^F(\tilde p)\Big]^R=  i\frac{\tilde p^2g^2}{4\pi{\mathcal{K}}|m_I^R|\upnu^2} -i\frac{\slashed{\tilde p} (\sign(m_I^R)+1)g^2}{4 \pi{\mathcal{K}}\upnu^2}+i\frac{|m_I^R| g^2}{4 \pi{\mathcal{K}}\upnu^2}.        \eea
\noindent \textcolor{black}{As a consistency check for this approximation, the momentum-dependent mass term implies the same relation to define the possible topological valley as in the case of the full self-energy expression presented in the motivating discussion in Eq. \eqref{ratio}. This point is going to be properly discussed in the next paragraphs.}\\ 
\indent  The inverse complete Feynman propagator reads
\begin{align}  \mathcal{S}^{-1}(\tilde p)=&-i\left[\slashed{\tilde p}-m_I^R-\Big[\Sigma_I^F(\tilde p)\Big]^R\right]\nonumber \\ =&-i\left[\slashed{\tilde p}\left( 1+\frac{g^2\big(\sign(m_I^R)+1\big)}{4\pi{\mathcal{K}} \upnu^2}\right)-m_I^R\left(1+\frac{ g^2\sign(m_I^R)}{4\pi {\mathcal{K}} \upnu^2} \right)-\frac{\tilde p^2}{4\pi {\mathcal{K}}\upnu^2|m_I^R| }      \right].     \end{align}
\indent According to Ref. \cite{wang1}, the object that encodes the topological properties of this kind of two-band system is the so-called topological Hamiltonian \cite{ger}, defined by 
\bea h_{top}=h_0+\Big[\gamma_0\Sigma_I^F(\vec p,E=0)\Big]^R=\sigma_i\eta_i^I(\vec p),\eea 
with $\sigma_i$ denoting the Pauli matrices and $h_0$ being the free Hamiltonian.\\
\indent The covariant formula of Ref. \cite{raya2} is then proved to be equivalent to the one given by \cite{wang2,volovik} 
\begin{equation}
{\mathcal{N}}_{Ch}^I=\frac{1}{4\pi} \int d^2k \ \hat{\eta}^I\cdot \big(\partial_{k_x}\hat{\eta}^I\times \partial_{k_y}\hat{\eta}^I\big),            \end{equation}
 with $\hat{\eta}^I=\eta^I/|\eta^I|$ and $\partial_{k_i}\equiv \frac{\partial}{\partial k_i}$.\\
\indent Regarding this vacuum-corrected fermionic response, considering a well-known result for a class of models including our system in vicinity of the on-shell regime, each valley is associated with the Chern number
\bea  {\mathcal{N}}_{Ch}^I=\frac{1}{2}\Bigg\{\sign\left[m_I^R\left(1+\frac{ g^2\sign(m_I^R)}{4\pi {\mathcal{K}} \upnu^2} \right)\right]+\sign\left[\frac{1}{4\pi {\mathcal{K}}|m_I^R|}\right] \Bigg\}. \label{tt1}     \eea
\noindent Considering the smallness of the radiative correction, one can infer from Eq. (\ref{tt1}) that one valley becomes topological and the other one keeps its trivial nature.

\section{On specific solutions for the renormalized theory}\label{sec6}
\subsection{Localized Boundary states}
\indent Regarding the renormalized fermionic structure, there is an interesting low-energy solution associated with the emergence of the momentum squared term. It can describe the effect of a hypothetical statistical interaction for graphene. The mentioned solution contains localized states with a unique drift velocity along a given line/boundary on the sample. It was originally applied in the study of the spin Hall effect in HgTe/CdTe quantum wells \cite{ber,shen2}. The necessary condition for its existence is a non-trivial Chern number implying a phenomenological signature for the valley asymmetry\footnote{For a statistically interacting graphene model, one renormalized valley becomes topological, whereas the other one does not.}. The present fermionic response enters as a special case of the dynamical system investigated in Ref. \cite{shen}, with an additional energy-dependent mass term. Accordingly, Refs. \cite{wang1,wang2} claim that although the topological Hamiltonian is the key object to encode the topology, the energy spectrum characterization should be given by the full renormalized structure. \\
\indent  We consider a semi-infinite planar sample with a boundary at $y=0$. Therefore, $p_x$ is an appropriate quantum number, and  $p_y$ is replaced by $-i\p_y$.  The renormalized system has the following general structure for the valleys
\bea \mathcal{A}E\Psi_I(p)=\Big(\tilde M_I \sigma_z-\mathcal{D}{\big( p_x^2-\p_y^2\big)}\sigma_z+\upnu \mathcal{A}\big(\sigma_x p_x-i\sigma_y\p_y\big)\Big)\Psi_I(p),                       \eea
with $\vec \sigma=(\sigma_x,\sigma_y,\sigma_z)$ denoting the Pauli matrices, $\mathcal{D}\equiv \frac{g^2}{4\pi \mathcal{K}|m_I^R|}$, and $\mathcal{A}\equiv \Big( 1+\frac{g^2\big(\sign(m_I^R)+1\big)}{4\pi{\mathcal{K}} \upnu^2}\Big)$. 
 The energy-dependent mass term is defined by
\bea  \tilde M_I\equiv m_I^R\left(1+\frac{g^2}{4\pi {\cal{K}}\upnu^2 }\sign(m_I^R)\right)+\frac{g^2E^2}{4\pi {\cal{K}}|m_I^R|\upnu^2}. \eea
\noindent Thus, considering the following ansatz for a localized boundary state that vanishes for $y\to 0$ and $y \to \infty$, 
 \begin{align}   \Psi_I(p) &= \begin{bmatrix}
           c(p_x) \\
           d(p_x) \\
          \end{bmatrix} e^{-y\Lambda},   \end{align}            one gets $\Lambda_{1,2}^2=p_x^2+F\pm \sqrt{F^2-\frac{\tilde M^2_I-E^2}{\mathcal{D}^2}}$, with $F\equiv \big( \mathcal{A}^2\upnu^2-2\tilde M_I \mathcal{D}  \big)/2\mathcal{D}^2$. According to Refs. \cite{shen,shen2}, a relation for the energy can be obtained
\bea \mathcal{A}E=\tilde M_I-\mathcal{D}\Lambda_1\Lambda_2-\mathcal{D}(\Lambda_1+\Lambda_2)p_x-\mathcal{D}p_x^2.  \eea
\noindent A solution that vanishes in the boundary but is concentrated near it, being normalizable, must have the following form
 \begin{align}   \Psi_I(p) &= \begin{bmatrix}
           c(p_x) \\
           d(p_x) \\
          \end{bmatrix} \Big(e^{-y\Lambda_1}- e^{-y\Lambda_2}\Big).   \end{align}
          The solution exists if both $\Lambda_1$ and $\Lambda_2$ have positive real parts.  Therefore, considering our specific case,  the limit $p_x=0$ also implies $E=0$ and 
\bea \Lambda_1\Lambda_2= -\frac{m_I^R\left(1+\frac{g^2}{4\pi {\cal{K}}\upnu^2 }\sign(m_I^R)\right)}{\frac{g^2}{4\pi {\cal{K}}|m_I^R|}},                \eea
with  $\Lambda_1\Lambda_2>0$  being the condition for the existence of the localized boundary state. Interestingly, it is the same existence condition that arises for the model with just a purely spatial momentum squared term \cite{shen}. It is not a surprising result if one considers that this is the condition to ensure a non-trivial Chern number for the system and the fact that this is the same for both models \cite{wang1}. By our previous discussions, if $\mathcal{N}^I_{Ch}\neq 0$, the associated valley presents low-energy boundary states with group velocity
 \bea {\cal{V}}=\upnu \sign \left(\frac{ g^2\sign(m_I^R)}{4\pi {\mathcal{K}} \upnu^2} \right)  \sign \left( 1+\frac{g^2\big(\sign(m_I^R)+1\big)}{4\pi{\mathcal{K}} \upnu^2}\right). \eea
whose modulus is the same as the graphene drift velocity $\upnu\approx \frac{c}{300}$. However, the direction of the movement along the border is dictated by the gap size and the  CS coefficient. Then, this system implies different phenomenological signatures for each valley, meaning that this interaction can be used as a valley filter.\\
\indent Concretely, considering this specific quasi-particle regime, for the case $p_x\approx 0$ and $E\approx 0$, we have\footnote{\textcolor{black}{Converting from natural units to micrometers ($\mu m$)}.}
\bea \Lambda_{1,2}^{-1}\sim 4  \times 10^{-1}\mu m   \eea
for the phase $|m_I|\approx 0.05$ meV,  $n\approx 100$ and $\upnu \approx \frac{1}{300}$. It defines the typical distribution of the boundary states. It reveals that these states are indeed tightly confined in the boundary. \textcolor{black}{ Then, increasing the natural number $n$, the topological states become even more confined. Interestingly, the coupling $g$, at this first approximation, has no contribution to this typical length. This is a consequence of the gauge symmetry requirements from the initial sections. Regarding the scales of established physical realizations, one can mention Hall devices associated with quantum resistance metrology. This is a relevant prototype example since it belongs to a wider class of low-energy planar phenomena that also includes our model.  For this example, the sample sizes \cite{metrology1,metrology2} have  order $30\ \mu m \times 150\ \mu m$. Therefore, focusing on samples with scales of this order also in our specific case, the topological confinement effect outlined here can indeed define a relevant spatial separation between the trivial and non-trivial topological valleys. As mentioned in Sec. \ref{intro11}, this property can furnish an additional organizing principle in valleytronics.  }

\subsection{The renormalized Landau levels}

\begin{figure}[H]
    \centering
    \includegraphics[width=14.8cm]{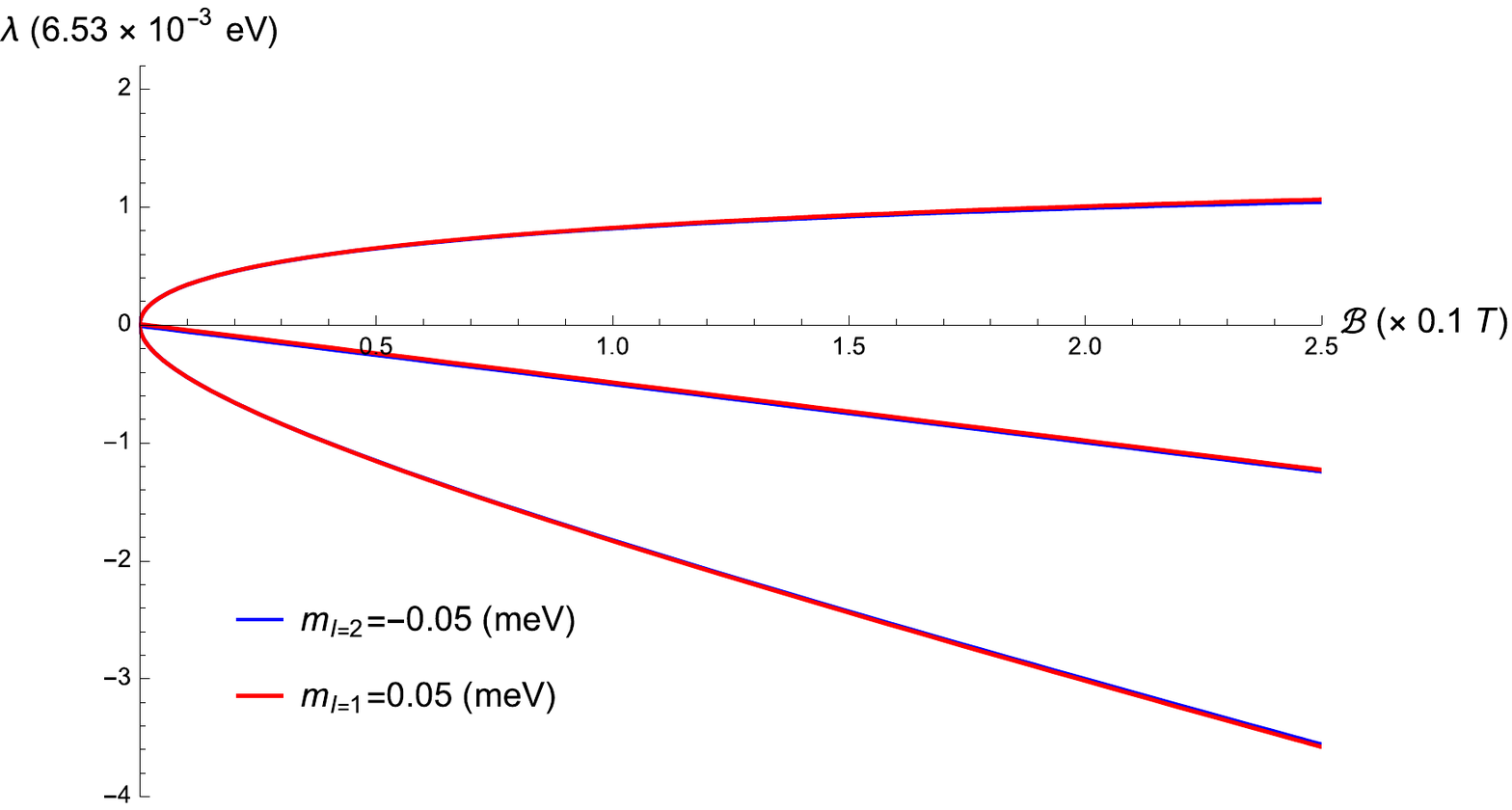}
    \caption{Dispersion of Landau levels. }
    \label{fig:1}
\end{figure}

\indent  Another signature of valley asymmetry due to the interaction with the CS model occurs for the renormalized Landau levels. In this case, the target configuration is a conventional graphene layer, renormalized by the intrinsic  topological interaction
in the presence of an external low magnetic $\mathcal{B}$, compatible to the continuum quasi-particle approximation. After performing the Peierls substitution to include the magnetic field effect,  $p_i\mapsto  p_i+eA_i(p)\equiv \Pi_i(p)$, with $A_i(p)$ denoting the electromagnetic potential, the following renormalized equation is achieved:
\bea {\mathcal{E}}\Psi_I={\mathcal{O}}\Psi_I, \eea
where \bea {\mathcal{E}}=\left(\begin{array}{ccccccc}
	\mathcal{A}E+\frac{g^2}{4\pi \mathcal{K}|m_I^R|\upnu^2}E^2&   0  \\
	  0 & \mathcal{A}E-\frac{g^2}{4\pi \mathcal{K}|m_I^R|\upnu^2}E^2     \\
	\end{array}\right).     \eea
\indent The independent Coulomb contribution for the self-energy \cite{coulomb}  can be accounted as another term entering the operator ${\mathcal{O}}$, to establish a more complete investigation. The effect of the Coulomb contribution is encoded by the following extra Hamiltonian term
\bea  \frac{27t'a^2}{2}\Pi_i(p)\Pi_i(p)\begin{bmatrix}
           1& 0\\
           0 & 1 \\
          \end{bmatrix},\eea
    with $\frac{a}{\sqrt{3}}\sim 0.142 $ nm and $t'\approx 0.4$ eV. Considering the noncommutative nature of  $\Pi_i(p)$, namely, 
\bea  \Big[\Pi_x,\Pi_y \Big ]=-ie\mathcal{B},       \eea
 a set of creation and annihilation operators can be employed to yield
\bea \Pi_x=\frac{e\mathcal{B}}{\sqrt{2}}(a+a^\dagger)\ , \qquad\qquad \Pi_y=-\frac{e\mathcal{B}}{\sqrt{2}i}(-a+a^\dagger), \eea
where the $a^\dagger$ denotes the creation operator associated to $\Pi_i(p)$,  with $[a,a^\dagger]=1$. 
 Therefore the operator ${\mathcal{O}}$  acquires the form
\bea {\mathcal{O}}=\left(\begin{array}{ccccccc}
	(c-d)\left(\mathfrak{n}-\frac{1}{2}\right)-{m}_I'& \upnu \sqrt{e{\mathcal{B}}}\sqrt{\mathfrak{n}}{\mathcal{A}}&     \\
	  \upnu \sqrt{e{\mathcal{B}}}\sqrt{\mathfrak{n}}{\mathcal{A}} &  (c+d)\left(\mathfrak{n}+\frac{1}{2}\right)+{m}_I'&    \\
	\end{array}\right),                                 \eea
 when acting on a class of eigenspinors 
 \bea \Psi_\mathfrak{n}^I= a_\mathfrak{n}\begin{bmatrix}
           |\mathfrak{n}-1\rangle \\
           0 \\
          \end{bmatrix}  +b_\mathfrak{n}\begin{bmatrix}
           0 \\
           |\mathfrak{n}\rangle \\
          \end{bmatrix}, \eea
which are associated with an equation for the generalized Landau energy levels for $\mathfrak{n}\geq 1$. Here, $d=-\frac{eg^2\mathcal{B}}{4\pi {\mathcal{K}}|m_I^R|}$, $c=\frac{27t'a^2e\mathcal{B}}{2}$ and $m_I'=m_I^R\left(1+\frac{g^2\sign(m_I^R)}{4\pi {\mathcal{K}}\upnu^2}  \right)$. Each component is related to the Fock states $\hat{\mathfrak{n}}|\mathfrak{n}\rangle=\mathfrak{n}|\mathfrak{n}\rangle$, with $\hat{\mathfrak{n}}=a^\dagger a$. 
\indent    The eigenvalues of the operator $\mathcal{O}$ are the following 
\bea  \uplambda_{\mathfrak{n}}^{(I\ \pm)}=\frac{(d+2\mathfrak{n}c)}{2}\pm \sqrt{ \upnu^2e{\mathcal{B}}\mathfrak{n}{\mathcal{A}}^2+\Big(d\mathfrak{n}+\frac{c}{2} +{ m}_I'\Big)^2},                 \eea
with $\mathfrak{n}\geq 1$ denoting the Landau levels, not to be confused with the CS coefficient depending on the parameter $n$, which is also a natural number. For $\mathfrak{n}=0$, the eigenstate 
 \bea \Psi_\mathfrak{n=0}^I= \begin{bmatrix}
           0 \\
           |\mathfrak{0}\rangle \\
          \end{bmatrix} \eea
has associated eigenvalue  given by 
\bea \uplambda_0^I=\frac{d+c+2m_I'}{2}.\eea
\indent For the case of interacting graphene on a magnetic field with strength $\approx 0.1$ T, one obtains $m_I^R\sim 0.05$ meV, $c\approx 1.55\times 10^{-5}$ eV, $d\approx -6.53 \times 10^{-3}$ eV, 
 ${\mathcal{A}}=1-\frac{1}{200}\big(\sign(m_I^R)-1\big)$, and $m_I'=5\times 10^{-5}\left(1+5\times 10^{-3}\sign(m_I)   \right)$ eV. The solutions for the energy spectrum can be properly chosen to yield real results for each value of the magnetic field, considering the setting $\mathcal{B}>0$ and $\mathcal{K}>0$. They are explicitly displayed as
\bea   E^{(1)I}_\mathfrak{n}=\frac{\mathcal{A}-\sqrt{\mathcal{A}^2-4\frac{\mathcal{D}}{\upnu^2}\uplambda^{(I-)}_\mathfrak{n}} }{2\frac{\mathcal{D}}{\upnu^2}}, \eea
including also the $\mathfrak{n}=0$ state. 
 There is also the solution
\bea   E^{(2)I}_\mathfrak{n}=\frac{-\mathcal{A}+\sqrt{\mathcal{A}^2+4\frac{\mathcal{D}}{\upnu^2}\uplambda^{(I+)}_\mathfrak{n}} }{2\frac{\mathcal{D}}{\upnu^2}}. \eea
\noindent The variable $\mathcal{D}$ was defined in the previous subsection describing the localized boundary solutions. The functional form is such that for the limit $\mathcal{D}\to 0$, which corresponds to no momentum-depending mass term, an analogous expression to the Landau levels of \cite{coulomb} is recovered.\\
\indent The valley symmetry breaking on the energy spectrum is  displayed in Fig. \ref{fig:1}.
\noindent It illustrates the behavior of some $\uplambda^{I\pm}_\mathfrak{n}$ associated to the first two Landau levels $\mathfrak{n}=0$, represented by two straight lines. The case $\mathfrak{n}=1$ regards 
 four curves corresponding to each valley denoted by $I=1,2$ and the solutions labeled by $\pm$. The ones with the ``$-$'' label are associated with negative eigenvalues. As one can easily verify, the valley symmetry breaking for $\mathfrak{n}=1$ levels becomes more evident when the magnetic field increases. For graphene-like materials such as germanene and borophene, with bigger mass gaps, the valley symmetry breaking becomes more prominent.

\section{Concluding remarks} \label{sec7}
\indent The effect of a statistical interaction for quasi-particles associated with Dirac-like materials was investigated in this paper. The renormalized quasi-particle structure presents a momentum-dependent mass term with topological implications. We proved that it recovers the same qualitative topological properties as a given phase of the Haldane-like models in \blt{the second-order expansion around
Dirac points of the reciprocal lattice}. The main motivation for this study relies on the recent advances in material engineering that can provide experimental realizations of such scenarios. The low energy cost associated with this kind of gauge field interaction and the possibility of \blt{anyonic} statistics for the planar quasi-particles were also important motivations. Considering low external magnetic fields, the effect on the vacuum-corrected renormalized graphene sample was carefully studied. The generalized Landau levels for statistically interacting graphene reveal an asymmetry between the valleys, comprising a phenomenological signature of the interaction with the CS mediator field. A solution for graphene materials interacting via an emerging topological field was also analyzed. Its existence condition depends on the topological nature of a given valley. Since the radiative corrections imply different natures for each one of them, just one can present the localized support and unique drift velocity. \\
\indent The observable properties strongly depend on the renormalized vacuum structure. The analysis of the topological gauge field in this scenario was provided by the KON indefinite metric quantization. This kind of quantization is demanded by the Poincaré structure associated with the gauge field. The \blt{Lorentz symmetry breaking scenario}, evincing quasi-particle drift velocity $\upnu <1$, was also discussed. A complete analysis of the \blt{fermionic two-point correlator} was obtained, considering both the perturbative and the non-perturbative regimes. A useful integral representation to derive the radiative correction was also furnished. As discussed, numerical estimates and considerations regarding the microscopic lattice structure suggested a perturbative nature for the class of systems studied here. 
In addition to the experimental research associated with Coulomb interaction for graphene quasi-particles, \cite{coulomb}, the interaction here discussed emulates the next-nearest neighbor hopping characteristic of the Haldane lattice. In this case, it is associated with an intrinsic generalized magnetic flux, as the interaction with the auxiliary gauge field indicates.\\
\indent \blt{The investigation of the three-point correlator for the system outlined here defines a natural next step as a relevant perspective to be implemented}. The associated advances in field theory tools are also of interest. Another possibility is the study of three-dimensional topological insulators employing a correlated field theory approach. Ref. \cite{Bazeia:2016est} can be useful in this context. Moreover, the correct introduction of boundaries to describe this system leads to improvements in the modeling of a wide range of analogous condensed matter phenomena. Finally, in a complementary formulation, regarding the charge-neutrality point, graphene can form the Dirac fluid. This new kind of strange metal is a relativistic plasma, consisting of strongly-interacting 
hole-electron pairs can be described by the relativistic Navier--Stokes equations in a disordered medium. 
The recent introduction of the Navier--Stokes equations with soft-hair in Ref. \cite{Ferreira-Martins:2021cga}
 may emulate some of the results heretofore obtained. Besides, since the Dirac fluid can be described by an AdS black brane as the holographic dual object to graphene at finite temperature in AdS/CMT, generalized four-dimensional black branes, also with 1-loop corrections, can play a prominent role in studying valley asymmetry \cite{Ferreira-Martins:2019wym,Kuntz:2019omq}. \\ \indent \textcolor{black}{Another perspective relies on the consideration of a wider set of interactions. In order to achieve a more realistic description of the scenario investigated here, one can include finite size effects, temperature, and interaction with impurities/external fields, among other backgrounds. These may induce effective hopping terms influencing the properties of the lattice. Moreover, according to \cite{refe1,refe2}, including more interactions can break a given class of topological order in Dirac materials. These complementary discussions define a natural next step in refining our model.}

\appendix

\section{On the generalized Hubbard--Stratonovich transformation}

\indent This appendix is devoted to proving that our procedure is indeed a generalization of the Hubbard--Stratonovich transformation. Although the auxiliary vector field has a non-trivial propagator, if one adds the correct ghost sector and the corresponding KON $B$ field, these extra fields do not contribute to the path integral, as expected. More explicitly, we are referring to the following transformation
\begin{align}  Z&=\int \mathcal{D}\mu_{\Psi}\exp\Bigg[i\int d^4x\Bigg( \frac{(J_\mu^1+J_\mu^2)\epsilon^{\mu \nu \alpha}\tilde{\p}_\nu (J_\alpha^1+J_\alpha^2)}{\mathcal{K}\tilde \Box} \Bigg) \Bigg]\nonumber\\&=N\!\int\! {\cal{D}}\mathcal{C}_{\mu}\mathcal{D}c\mathcal{D}\bar c\mathcal{D}B\mathcal{D}\mu_{\Psi}\exp\Bigg[i\!\int d^4x\Big(  \frac{\mathcal{K}}{2}\mathcal{C}_\mu \epsilon^{\mu \nu \alpha}\tilde \p_\nu \mathcal{C}_\alpha+ \sum_{I=1}^2 \Bar{\Psi}_I\gamma^\mu \Psi_I\mathcal{C}_\mu+B\tilde{\p}_\mu \mathcal{C}^\mu+\frac{\alpha}{2} B^2+\bar c\tilde \Box c\Big) \Bigg]         \end{align}
         with $J^\mu_I(x)=\Bar{\Psi}_I(x)\gamma^\mu \Psi_I$, $N$ being a non-physical normalization constant,  and $\mathcal{D}\mu_{\Psi}$ denoting the fermionic measure.  Although it is not relevant for the $T=0$ discussion, finite temperature effects are strongly dependent on differential operator normalization. Therefore, it is worth mentioning that this process does not add any new degrees of freedom. It is indeed a generalized Hubbard-Stratonovich transformation.

\section { The calculation of Dirac brackets}

\indent Here the Dirac brackets associated with the equal time commutators are obtained through the correspondence principle. The $B$ field formalism fixes the action in such a way that there are no first-class constraints, meaning that the Dirac brackets can be considered with all the constraints in the strong form. It can be straightforwardly understood by considering the matrix of all the Poisson brackets for the bosonic sector of the  constraints
\bea \mathcal{G}^{IJ}(x,y)=\{\Phi^I(x),\Phi^J(y)\}=\left(\begin{array}{ccccc}
	\epsilon^{nk}{\cal{K}}& 0& 0&       \\
	0 & 0 & 1 &        \\
	0& -1 & 0        \\  
\end{array}\right)\delta^2(x-y),  \eea

with \bea\Phi^I(x)=\left(\begin{array}{cccccc}
	\pi_i(x)+\frac{\epsilon_{ij}A^j(x){\cal{K} }}{2}& \\
	\pi_B(x) &     \\
	\pi_0(x)-B(x)&      \\  
\end{array}\right).\eea
\indent  the inverse matrix is given by
\bea \tilde{\mathcal{G}}^{IJ}(x,y)=\{\Phi^I(x),\Phi^J(y)\}=\left(\begin{array}{ccccc}
	\frac{\epsilon_{nk}}{\cal{K}}& 0& 0&  \\
	0 & 0 & 1 &    \\
	0& -1 & 0    \\   \end{array}\right)\delta^2(x-y). \eea
\indent Some of the brackets projected on the reduced phase space are expressed as 
\begin{multline} \{A_i(x),\pi_j(y)\}_D=\{A_i(x),\pi_j(y)\}-\int d^2w d^2z\{A_i(x),\Phi^1_n(w)   \}\frac{\epsilon_{mn}}{{\cal{K}}}\{\Phi^1_m(z),\pi_j(y)   \}=\frac{\delta_{ij}}{2}\delta^2(x-y),\end{multline}
instead of $\delta_{ij}\delta^2(x-y)$. It permits us to derive the correct factor for the commutator between the spatial vector fields. Also, 
 $\pi_B(x)$ disappears from the dynamics and $\pi_0(x)=B(x)$ is valid in the strong form. An analogous procedure for the fermionic sector ensures that the associated momenta constraints can be considered in the strong form.

\section{Definition of Pauli--Jordan and double pole distributions}

\indent The Pauli--Jordan function is an important distribution to define the physical field commutators.  It obeys a Klein-Gordon equation and has the following definition
\begin{align}&iD(x-y, s)=D^+(x-y, s)+D^-(x-y, s), \qquad\quad \tilde \Box D(x-y,s)=-sD(x-y,s),\nonumber\\&  D(x-y,s)|_{x_0=y_0}=0,\qquad\qquad\qquad\quad \p_0D(x-y,s)|_{x_0=y_0}=-\delta^2(x-y), \nonumber\\  &  D^\pm(x-y, s)=\mp \frac{1}{(2\pi)^2}\int d^3p\ \delta(\tilde p^2-s)\Theta(\pm p_0)e^{-ip.(x-y)},
\end{align}
in terms of its initial conditions. It can be expressed as a sum of positive and negative frequency functions.\\
\indent The time-ordered, or Feynman, distribution reads
\bea     D_F(x-y,m^2)=i\int \frac{d^3p}{(2\pi)^3}\frac{e^{-ip.(x-y)}}{(\tilde p^2-m^2+i\sigma)}          \eea
with $\sigma\to 0$.\\
\indent The so-called double pole distribution and its initial conditions are defined below
\begin{align} &(\tilde \Box+s)E(x-y,s)=D(x-y,s), \qquad\qquad \p_0^3E(x-y,s)|_{x_0=y_0}=-\delta^2(x-y), \nonumber \\
&E(x-y,s)|_{x_0=y_0}=0,\qquad\quad 
E(x,m^2_I)\equiv-\int d^3u\ \varepsilon(x,0,u)D(x-u,m^2_I)D(u,m^2_I),  \end{align}
in terms of the Pauli--Jordan functions.\\
\indent The commutator version of the bosonic self-energy is depicted as follows
 \bea g^2\langle 0 |\Big[J_\mu(x),J_\nu(y)\Big]|0\rangle =
  \sum_I g^2tr\Big(\gamma_\mu S^+_I(x-y)\gamma_\nu S^-_I(y-x)-\gamma_\mu S^-_I(x-y)\gamma_\nu S^+_I(y-x)       \Big),                    \eea
  in its first-order approximation in terms of the free fermion distributions.\\
 \indent The  Fourier transform of this object
\begin{align} g^2\langle 0 |\Big[J_\mu(\tilde k),J_\nu(-\tilde k)\Big]|0\rangle=
   &-\sum_I\frac{g^2}{8\upnu^2}\Big(1+\frac{4m^2_I}{\tilde k^2}\Big)\theta(\tilde k^2-4m^2_I)\frac{{\rm sign}(k_0)}{\sqrt{\tilde k^2}}\Big(\tilde k_\mu\tilde k_\nu-\tilde k^2\eta_{\mu \nu}\Big)\nonumber \\&+\sum_I\frac{ig^2m_I}{2\upnu^2}\epsilon_{\mu \nu \alpha}\tilde k^\alpha \frac{1}{\sqrt{\tilde k^2}}\theta(\tilde k^2-4m^2_I)\,{\rm sign}(k_0)  \end{align}
is closely related to the spectral density of the system. Employing the latter, the causal, the retarded/advanced and the Feynman distributions can be obtained in a spectral representation.

\section{Gamma matrices algebra}

\indent The two-dimensional faithful representation for fermions is given in terms of Pauli matrices with $\gamma^0=\sigma_3$, $\gamma^1=i\sigma_1$ and $ \gamma^2=i\sigma_2$, 
 obeying
\bea \Big\{\gamma^\mu,\gamma^\nu  \Big\}=2I\eta^{\mu \nu },\eea
where $\gamma^\mu\gamma^\nu=\eta^{\mu \nu }-i\epsilon^{\mu \nu \alpha }\gamma_\alpha$, for $I$ denoting the $2\times 2$ unity matrix.  There are the following identities for the traces
\begin{align} tr(\gamma^\mu)=& 0 \ , \ tr(\gamma^\mu\gamma^\nu)=2\eta^{\mu \nu },\nonumber \\
tr(\gamma^\mu\gamma^\nu\gamma^\rho)=&-2i\epsilon^{\mu \nu \rho },\nonumber \\
tr(\gamma^\mu\gamma^\nu\gamma^\rho \gamma^\sigma)=&2\Big(\eta^{\mu \nu}\eta^{\rho \sigma}-\eta^{\mu \rho}\eta^{\nu \sigma}+\eta^{\mu \sigma}\eta^{\nu \rho}   \Big).
\end{align}

\acknowledgments
\indent G.B.G. thanks the São Paulo Research Foundation -- FAPESP  Post Doctoral grant No. 2021/12126-5. B.M.P. thanks CNPq for partial support. R.d.R.~is grateful to The S\~ao Paulo Research Foundation FAPESP (Grant No. 2021/01089-1 and No. 2022/01734-7), the National Council for Scientific and Technological Development -- CNPq (Grant No. 303390/2019-0), and the Coordination for the Improvement of Higher Education Personnel (CAPES-PrInt  88887.897177/2023-00), for partial financial support.

\end{document}